\documentclass[onecolumn,preprintnumbers,eqsecnum,superscriptaddress,12pt]{revtex4}
\usepackage{amsfonts}
\usepackage{amssymb}
\usepackage{amsmath}
\usepackage{epsfig}
\usepackage{graphicx}
\usepackage{dcolumn}
\usepackage{bm}

\usepackage{color}

\newcommand{\beq}{\begin{equation}}
\newcommand{\eeq}{\end{equation}}
\newcommand{\beqs}{\begin{eqnarray}}
\newcommand{\eeqs}{\end{eqnarray}}

\newcommand{\be}{\begin{equation}}
\newcommand{\ee}{\end{equation}}
\newcommand{\bea}{\begin{eqnarray}}
\newcommand{\eea}{\end{eqnarray}}

\begin{document}

\title{Perihelion precession and deflection of light in the general spherically symmetric spacetime}

\author{Ya-Peng Hu}\email{huyp@nuaa.edu.cn}
\address{College of Science, Nanjing University of Aeronautics and Astronautics,
Nanjing 210016, China}
\address{INPAC, Department of Physics, and Shanghai Key Laboratory of Particle Physics and Cosmology, Shanghai Jiao Tong University, Shanghai 200240, China }

\author{Hongsheng Zhang}\email{hongsheng@shnu.edu.cn}
\address{Shanghai United Center for Astrophysics (SUCA), Shanghai Normal University, 100 Guilin Road, Shanghai 200234, China}

\author{Jun-Peng Hou}\email{jack.hjp@gmail.com}
\address{College of Science, Nanjing University of Aeronautics and Astronautics,
Nanjing 210016, China}

\author{Liang-Zun Tang}\email{tangliangzun1234@163.com}
\address{College of Science, Nanjing University of Aeronautics and Astronautics,
Nanjing 210016, China}

\begin{abstract}
In this paper, the perihelion precession and deflection of light have been investigated in the $4$-dimensional general spherically symmetric spacetime, and the master equation is obtained. As the application of this master equation, the Reissner-Nordstorm-AdS solution and
Clifton-Barrow solution in $f(R)$ gravity have been taken for the examples. We find that both the electric charge and $f(R)$ gravity can affect the perihelion precession and deflection of light, while the cosmological constant can only effect the perihelion precession. Moreover, we clarify a subtlety in the deflection of light in the solar system that the possible sun's electric charge is usually used to interpret the gap between the experiment data and theoretical result. However, after also considering the effect from the sun's same electric charge on the perihelion precession of Mercury, we can find that it is not the truth.

\end{abstract}

\maketitle


\newpage

\section{Introduction}
In the history, the perihelion precession of Mercury and deflection of light in the solar system are two well-known phenomena to
check the correctness of general
relativity \citep{weinberg72,Misner:1974qy,Wald:1984rg,Turyshev:2007qy,Turyshev:2008dr}.
Nowadays, we know that the foundation of general relativity is a
very significant event in modern physics. Not only general
relativity can give new insights into our understanding of gravity,
but also it has been the basic theory in our modern
cosmology~\citep{weinberg72}. Since the perihelion
precession and deflection of light are usually constrained in the
solar system or some planet such as
Mercury~\citep{weinberg72,Misner:1974qy,Wald:1984rg}. Therefore, it
will be worthy to investigate the perihelion precession and
deflection of light in the more general case. In this paper, the
perihelion precession and deflection of light have been considered in the $4$-dimensional general spherically symmetric spacetime. Since the perihelion
precession and deflection of light can be treated as the time-like and null
geodesic in spacetime~\citep{Wald:1984rg}, thus we obtain the corresponding main
equation in the $4$-dimensional general spherically symmetric spacetime.

Note that, due to the Birkhoff theorem or the generalization of the Birkhoff theorem in Einstein gravity, the general spherically symmetric spacetimes in Einstein gravity are very limited. However, the Birkhoff theorem can be invalid in some modified gravities, i.e. $f(R)$ gravity which is a kind of higher derivative gravity theory~\cite{Sotiriou:2008rp,Cai:2009qf}. Therefore, as the application of the master equation in the $4$-dimensional general spherically symmetric spacetime, we take the Reissner-Nordstorm-AdS solution in Einstein gravity and Clifton-Barrow solution in $f(R)$ gravity~\cite{CB,Zhang:2014goa} for the two examples to discuss the corresponding perihelion precession and deflection of light. For the Reissner-Nordstorm-AdS solution with the electric charge and cosmological constant, we find that the electric charge can affect both the perihelion precession and deflection of light, while the cosmological constant can only affect the perihelion precession, which are consistent with the results in the previous work~\citep{Peebles:2002gy,Weinberg:1988cp,Albrecht:1999rm,Li:2011sd,Kagramanova:2006ax}. It should be emphasized that there is a subtlety in the previous work to discuss the well-known deflection of light in the solar system. The subtlety is that the possible sun's electric charge is usually used to interpret the gap between the experiment data and theoretical result in the deflection of light in the solar system, because the electric charge can affect the deflection of light~\citep{Jetzer:2006gn,Iorio:2007ub,Iorio:2011p,Iorio:2011aa,Sereno:2006mw,Sereno:2003nd,Sereno:2006re,Sereno:2007rm,Sereno:2008kk}. However, after using our results and also considering the effect from the sun's same electric charge on the perihelion precession of Mercury~\citep{Dyson-Eddington-Davidson-1920,Kennefick:2009}, we can find that it is not the truth, i.e. the gap between the experiment data and
theoretical result for the deflection of light in solar system can not completely come from the possible sun's electric charge. For the Clifton-Barrow solution in $f(R)$ gravity~\cite{CB,Zhang:2014goa}, there is a parameter $d$ which is the power of Ricci scalar $R$ in one kind of $f(R)$ gravity in this solution. It is obvious that the Birkhoff theorem is invalid for this kind of $f(R)$ gravity, since the Schwarzschild metric is also an another static vacuum solution. For the simplicity and making some explicit comparison with other results, we just consider the case with small $d$ in our paper, because $d=0$ is just the Einstein gravity in our setting. Therefore, the parameter $d$ can represent the derivation of $f(R)$ gravity to Einstein gravity. From our results, we can easily find that the $f(R)$ gravity can both effect the perihelion precession and deflection of light.

The rest of the paper is organized as follows. In Sec.~II, after investigating the perihelion precession and
deflection of light in the $4$-dimensional general spherically symmetric spacetime, the master equation is obtained. In Sec.~III, the Reissner-Nordstorm-AdS solution in Einstein gravity and Clifton-Barrow solution in $f(R)$ gravity are taken as the two examples for the application of master equation, and the effects from the cosmological constant, electric charge and $f(R)$ gravity on the perihelion precession and deflection of light are investigated. Finally, besides a simple conclusion, we also make several discussions according to the experiment data in Sec.~IV.

\section{perihelion precession and deflection of light in the $4$-dimensional spherically symmetric spacetime: general case}
For the $4$-dimensional general spherically symmetric spacetime, its line element can be
\begin{equation}
ds^2=-f(r)dt^2+\frac{dr^2}{h(r)}+r^2(d\theta^2+\text{sin}^2\theta d \varphi^2), \label{Metric}
\end{equation}
Since the perihelion precession and deflection of light are usually treated as the time-like and null geodesic in spacetime, respectively. Therefore, we first discuss the geodesics $\gamma(\tau)$ in the above general spherically symmetric spacetime. We set the geodesic $\gamma(\tau)$ expressed in the above coordinates $x^{\mu}=(t,r,\theta,\varphi)$ as $x^{\mu}(\tau)$, which are satisfied
\begin{equation}
\frac{d^2x^{\mu}}{d\tau^2}+\Gamma^{\mu}_{\nu\sigma}\frac{dx^{\nu}}{d\tau}\frac{dx^{\sigma}}{d\tau}=0.
\end{equation}
Generally, after the above equation is solved, then the geodesic $\gamma(\tau)$ is obtained. However, considering the symmetry of spacetime (\ref{Metric}), we could find in the following that there is a more simple way to obtain the geodesic $\gamma(\tau)$. First, we can find that one component of the geodesic $\gamma(\tau)$ can always be chosen as $\theta (\tau)= \pi/2$, which means that the geodesic can always be chosen to lay in the equatorial plane of the spherically symmetric spacetime. Therefore, the geodesic can be simplified
\begin{equation}
t=t(\tau),~r=r(\tau),~\theta=\pi/2,~\varphi=\varphi(\tau).
\end{equation}
If we let $U^a\equiv(\frac{\partial}{\partial\tau})^a$ to be the tangent vector of geodesic $\gamma(\tau)$, we could define
\begin{equation}
\kappa=-g_{ab}U^aU^b=-g_{ab}(\frac{\partial}{\partial\tau})^a(\frac{\partial}{\partial\tau})^b,
\end{equation}
Thus $\kappa=1$ corresponds to the time-like geodesic, while $\kappa=0$ is the null geodesic.
After inserting $(\frac{\partial}{\partial\tau})^a=\frac{dx^{\mu}}{d\tau}(\frac{\partial}{\partial x^{\mu}})^a$ and the metric (\ref{Metric}), we could obtain
\begin{equation}
-\kappa=g_{ab}(\frac{\partial}{\partial\tau})^a(\frac{\partial}{\partial\tau})^b=-f(r)(\frac{dt}{d\tau})^2
+h(r)^{-1}(\frac{dr}{d\tau})^2+r^2(\frac{d\varphi}{d\tau})^2.
\label{MasterEq}
\end{equation}
where we have used $\theta=\pi/2$. Second, note that $(\partial/\partial t)^a$ and $(\partial/\partial \varphi)^a$ are two killing vectors in the spherically symmetric spacetime (\ref{Metric}). Therefore, there are two conserved quantities along the geodesic $\gamma(\tau)$
\begin{eqnarray}
E&=&-g_{ab}(\frac{\partial}{\partial t})^a(\frac{\partial}{\partial \tau})^b=f(r)\frac{dt}{d\tau}, \label{ConservedE}\\
L&=&g_{ab}(\frac{\partial}{\partial
\varphi})^a(\frac{\partial}{\partial
\tau})^b=r^2\frac{d\varphi}{d\tau}. \label{ConservedL}
\end{eqnarray}
where the physical meanings of $E$ and $L$ can be found in detail in Ref.~\cite{Wald:1984rg}, i.e., $E$ can be interpreted as the total energy (including gravitational potential energy) per unit rest mass of a particle in the timelike case, while $L$ can be interpreted as the angular momentum per unit rest mass of a particle. In addition, $\hbar E$ and $\hbar L$ can be interpreted as the total energy and angular momentum of a photon in the null case, respectively.

After inserting (\ref{ConservedE}) and (\ref{ConservedL}) into (\ref{MasterEq}), we could obtain
\begin{equation}
(\frac{dr}{d\tau})^2=\frac{h(r)}{f(r)}E^2-h(r)(\kappa+\frac{L^2}{r^2}).
\label{Eqrtau}
\end{equation}
Obviously, the above equation contains only one function $r(\tau)$, which could be solved in principle. Hence, after inserting the solved $r(\tau)$ into (\ref{ConservedE}) and (\ref{ConservedL}), the rest components $t(\tau)$ and $\varphi(\tau)$ of geodesic could be finally obtained.

It should be pointed that perihelion precession and deflection of light are usually related to the orbit of geodesic, i.e. $r(\varphi)$. Therefore, it is convenient to change the equation (\ref{Eqrtau}) as
\begin{equation}
(\frac{dr}{d\varphi})^2(\frac{L}{r^2})^2=\frac{h(r)}{f(r)}E^2-h(r)(\kappa+\frac{L^2}{r^2}),
\label{MasterEq1}
\end{equation}
where we have used the equation (\ref{ConservedL}). In addition, it has been found that the coordinate $\mu\equiv1/r$ is more convenient than $r$ to discuss the perihelion precession and deflection of light. Thus the master equation investigated in our paper could be simply obtained from equation (\ref{MasterEq1}) by changing $r$ into $\mu$
\begin{equation}
(\frac{d\mu}{d\varphi})^2=\frac{h(\mu)}{f(\mu)}(\frac{E}{L})^2-h(\mu)(\frac{\kappa}{L^2}+\mu^2).
\label{MasterEq2}
\end{equation}

\section{perihelion precession and deflection of light in the $4$-dimensional spherically symmetric spacetime: special case}
Since functions $h(r)$ and $f(r)$ are usually different for each spherically symmetric spacetime, thus the perihelion precession and deflection of light may be different in different spherically symmetric spacetime. Therefore, we can use the differences in perihelion precession and deflection of light to extract the information in $h(r)$ and $f(r)$. In this section, we will take two spherically symmetric solutions for examples as the application of master equation (\ref{MasterEq2}) to discuss the corresponding perihelion precession and deflection of light. The first solution is the Reissner-Nordstorm-AdS solution, which is a well-known solution in Einstein gravity with the cosmological constant and electric charge. While the other solution is the Clifton-Barrow solution in $f(R)$ gravity. The $f(R)$ gravity is a kind of higher derivative gravity theory, and the Clifton-Barrow solution can be considered as a generalization of Schwarzschild solution in Einstein gravity. For this solution, the advantages are that not only it is the spherically symmetric spacetime in $f(R)$ gravity, but also $h(r)$ and $f(r)$ are different.

\subsection{Reissner-Nordstorm-AdS solution}
For the Reissner-Nordstorm-AdS solution, the two functions $h(r)$ and $f(r)$ are
\begin{equation}
f(r)=h(r)=1-\frac{2M}{r}+\frac{\Lambda}{3}r^2+\frac{Q^2}{r^2}. \label{M3_1}
\end{equation}
where $\Lambda$ is the cosmological constant, and $Q$ is the electric charge. Therefore, Eq.(\ref{MasterEq2}) becomes
\begin{equation}
(\frac{d\mu}{d\varphi})^2=(\frac{E}{L})^2-(1-2M\mu+\frac{\Lambda}{3\mu^2}+\mu^2 Q^2)(\frac{\kappa}{L^2}+\mu^2),
\label{MasterEqLambda}
\end{equation}
For the perihelion precession, one usually considers the time-like geodesic, i.e. $\kappa=1$. Therefore, Eq.(\ref{MasterEqLambda}) for the time-like geodesic can be
\begin{equation}
\frac{d^2\mu}{d\varphi^2}+(1+\frac{Q^2}{L^2})\mu=\frac{M}{L^2}+3\mu^2M+\frac{\Lambda}{3\mu^3L^2}-2Q^2\mu^3.
\label{MasterEqLambda1}
\end{equation}
Obviously, compared with the case in Newton's gravity $\frac{d^2\mu}{d\varphi^2}+\mu=\frac{M}{L^2}$,
the term $3\mu^2M$ comes from the correction of general relativity, while the
last two terms are from the cosmological
constant and electric charge, and the above equation can return to the well-known Schwarzschild case when $\Lambda=Q=0$. Note that, the analytical solution of (\ref{MasterEqLambda1}) is absent like the Schwarzschild case. However, there is an approximation solution of (\ref{MasterEqLambda1})
\begin{eqnarray}
\mu(\varphi)&=&\frac{M}{L^2}(1+e
\text{cos}\varphi)+\frac{3M^3}{L^4}(1+e\varphi \text{sin}\varphi+e^2
(\frac{1}{2}-\frac{1}{6}\text{cos}2\varphi))+\frac{\Lambda
L^4}{3 M^3}(1-\frac{3}{2}e \varphi \text{sin}\varphi)\notag\\
&-&\frac{MQ^2}{L^4}(1+\frac{1}{2}e\varphi \text{sin}\varphi)
-\frac{2Q^2M^3}{L^6}(1+\frac{3}{2}e\varphi \text{sin}\varphi+3 e^2
(\frac{1}{2}-\frac{1}{6}\text{cos}2\varphi)),
\label{PSolution}
\end{eqnarray}
in the following conditions
\begin{equation}
3 M \mu^2\ll\mu,~~\frac{\Lambda}{3\mu^3L^2}\ll\mu,~~\frac{Q^2}{L^2}\ll1,~~2Q^2\mu^2\ll1 \label{Condition}
\end{equation}
where $\mu(\varphi)=\frac{M}{L^2}(1+e \text{cos}\varphi)$ is the analytical elliptical solution which has already been found in Newton's gravity, and $e$ is the orbital eccentricity which has been considered as a small constant. Therefore, Eq.(\ref{PSolution}) can be further reduced after neglecting the high order terms
\begin{equation}
\mu({\varphi})=\frac{M}{L^2}
\{1+e[\text{cos}\varphi+(\frac{3M^2}{L^2}-\frac{\Lambda
L^6}{2M^4}-\frac{Q^2}{2L^2})\varphi \text{sin}\varphi] \},
\end{equation}
where the conditions $\frac{3 M^2}{L^2}<<1$, $\frac{\Lambda
L^6}{3M^4}<<1$ have also been assumed. Note that, the above equation could be further simplified as
\begin{equation}
\mu(\varphi)=\frac{M}{L^2}[1+ e \text{cos}(\varphi - \varepsilon
\varphi)],
\end{equation}
where we have set
\begin{equation}
\varepsilon = (\frac{3M^2}{L^2}-\frac{\Lambda L^6}{2M^4}-\frac{Q^2}{2L^2}).
\end{equation}
For the perihelion of orbit $r(\varphi)$, it satisfies $\text{cos}(\varphi - \varepsilon \varphi)=1$, and hence $\varphi=2 \pi + 2 \pi \varepsilon$.
Therefore, the precession angle of perihelion is
\begin{equation}
\Delta \varphi=2\pi \varepsilon=2\pi(\frac{3M^2}{L^2}-\frac{\Lambda
L^6}{2M^4}-\frac{Q^2}{2L^2}). \label{M3_7}
\end{equation}
Note that, here $L$ is the angular momentum per unit rest mass of the particle along the time-like geodesic. Obviously, both the cosmological constant and electric charge can affect the precession angle of perihelion. When both the cosmological constant and electric charge disappear in (\ref{M3_7}), the result recovers the standard general relativity result with Schwarzschild solution $\Delta \varphi=\frac{6\pi M^2}{L^2}\approx\frac{6\pi M}{a}$, where $a$ is the semimajor axis of the ellipse and $a\approx\frac{L^2}{M}$ when the eccentricity $e$ is small.

Next, we will discuss the deflection of light in the Reissner-Nordstorm-AdS spacetime. In this case, the corresponding geodesic is the null geodesic, i.e, $\kappa=0$. Similar with the procedure dealt with in the perihelion precession, we need find out the approximation solution of the orbit of deflection of light. The master equation is
\begin{equation}
(\frac{d\mu}{d\varphi})^2+\mu^2=\frac{E^2}{L^2}+2M\mu^3-\frac{\Lambda}{3}-Q^2\mu^4,
\label{Sequation}
\end{equation}
Note that, the cosmological constant term is just a constant term like $\frac{E^2}{L^2}$. Therefore, the cosmological constant will not affect the deflection angle of light. In fact, Eq.(\ref{Sequation}) can be further simplified as
\begin{equation}
\frac{d^2\mu}{d \varphi^2}+\mu=3M\mu^2-2Q^2\mu^3. \label{M_SDPer}
\end{equation}
which shows more clearly that the cosmological constant does not affect the deflection angle. Note that, the approximation solution of Eq.(\ref{M_SDPer}) is
\begin{equation}
\mu(\varphi)=\frac{1}{l} \text{sin}\varphi+\frac{M}{l^2}(1-
\text{cos}\varphi)^2-\frac{2Q^2}{l^3}(-\frac{3}{8}\varphi
\text{cos}\varphi+\frac{1}{32}\text{sin}3\varphi),
\end{equation}
where $l$ is a constant, $\mu(\varphi)=\frac{1}{l} \text{sin}\varphi$ is in fact a straight line expressed in polar coordinates $(\mu,\varphi)$, and we have used the condition $\mu(0)=0$. Therefore, the deflection angle of light $\beta$ can be obtained from the equation
\begin{equation}
\mu(\pi+\beta)=0.
\end{equation}
After using the approximation
\begin{equation}
\text{sin}(\pi+\beta) \approx -\beta,\text{cos}(\pi+\beta) \approx
-1,
\end{equation}
the angle is
\begin{equation}
\beta=\frac{4M}{l}-\frac{3\pi Q^2}{4l^2}. \label{QstarD}
\end{equation}
where the first term is just the well-known deflection angle of light in Schwarzschild spacetime, while the second term is from the affect of the electric charge.

\subsection{Clifton-Barrow solution}

In this subsection, we will investigate the perihelion precession and deflection of light in the Clifton-Barrow solution in $f(R)$ gravity. In a general $f(R)$ theory, the uniqueness theorem of the spherically symmetric space becomes invalid. $R$ may be not zero even for vacuum solutions.  Considering the 4 dimensional action in the following form,
   \begin{equation}
   S=\frac{1}{16\pi G }\left(\int_{\cal M} d^{4}x\sqrt{-{\rm det}(g)}~R^{d+1}+\int_{\cal \partial M} d^{3}x\sqrt{-{\rm det}(h)}~2B\right),
  \label{action1}
  \end{equation}
   where $d$ is a constant, $B$ represents the corresponding boundary term for the $f(R)$ term, $g$ is the $4$-dimensional metric, $h$ denotes the induced
metric on the boundary, the corresponding field equation reads,
    \begin{equation}
   (1+d)R^{d}R_{\mu \nu }-\frac{1}{2}R^{d+1}g_{\mu \nu }-(1+d)\nabla _{\mu }\nabla
_{\nu }R^{d}+g_{\mu \nu }(1+d)\square R^{d}=0.
\label{field}
  \end{equation}
  It is easy to see that the Schwarzschild metric is a solution for the above equation, since Schwarzschild metric has a vanishing Ricci scalar $R$. In addition, the $f(R)$ gravity permits a non-trivial spherically symmetric solution other than the Schwarzschild metric, and the called Clifton-Barrow solution also solves the field equation \cite{CB},
 \begin{equation}
ds^{2}=-f(r)dt^{2}+\frac{dr^{2}}{h(r)}+r^{2}(d\theta ^{2}+\sin ^{2}\theta
d\phi ^{2}),
\label{ansatzs}
\end{equation}
where
\begin{eqnarray}
f(r) &=& r^{2d\frac{1+2d}{1-d}} + \frac{c}{r^{\frac{1-4d}{1-d}}}, \notag\\
h(r) &=& \frac{(1-d)^2}{(1-2d+4d^2)(1-2d(1+d))}(1+\frac{c}{r^\frac{1-2d+4d^2}{1-d}}). \label{CB Solution}
\end{eqnarray}
 Here $c$ is a constant, which reduces to the Schwarzschild mass parameter $c=-2M$ in Einstein gravity.

Therefore, one can directly insert (\ref{CB Solution}) into the master equation (\ref{MasterEq2}) to discuss its perihelion precession and deflection of light.
Note that, after inserting the two functions above into (\ref{MasterEq2}), it will be found that the equations are difficult to be solved directly. In order to explicitly show the effects on perihelion precession and deflection of light from the $f(R)$ gravity, here we just consider the case that the constant $d$ is very small since $d=0$ is just the Einstein gravity. In this case, the master equation (\ref{MasterEq2}) turns out to be
\begin{eqnarray}
\frac{d^2\mu}{d\varphi^2}+\mu&=&\frac{\kappa M}{L^2}+3M\mu^2+[\frac{\kappa M}{L^2} + \frac{E^2}{L^2}\frac{1}{\mu}-2\mu+5M\mu^2+(\frac{\kappa M}{L^2}+3M\mu^2)\text{ln}\frac{1}{\mu}]d,
\end{eqnarray}
where we have kept the linear terms of $d$ and neglected the higher order terms.

For the perihelion precession, i.e. $\kappa=1$, the above equation is still a little complicated, which can be further simplified in the larger radius case, i.e. $\mu=1/r\sim0$. Thus, the above equation is
\begin{eqnarray}
\frac{d^2\mu}{d\varphi^2}+\mu&=&\frac{\kappa M}{L^2}+3M\mu^2+ \frac{E^2}{L^2}\frac{d}{\mu},
\end{eqnarray}
and the approximate solution can be easily obtained
\begin{eqnarray}
\mu(\varphi)&=&\frac{M}{L^2}(1+e \text{cos}\varphi)+\frac{3M^3}{L^4}(1+e\varphi \text{sin}\varphi+e^2 (\frac{1}{2}-\frac{1}{6}\text{cos}2\varphi))\notag\\
&&+\frac{E^2}{M}(1-\frac{1}{2}e\varphi\text{sin}\varphi)d,
\end{eqnarray}
which can be further simplified
\begin{equation}
\mu(\varphi)=\frac{M}{L^2}\{
1+e[\text{cos}\varphi+(\frac{3M^2}{L^2}-\frac{E^2L^2}{2M^2}d)\varphi
\text{sin}\varphi] \}.
\end{equation}
Therefore, the angel of perihelion precession is
\begin{equation}
\Delta \varphi=2\pi
\varepsilon=2\pi(\frac{3M^2}{L^2}-\frac{E^2L^2}{2M^2}d).
\end{equation}
From the above simple discussion, we can easily find that there can be indeed effect from the $f(R)$ gravity on the perihelion precession.

For the deflection of light, i.e.$ \kappa=0$, the master equation becomes
\begin{equation}
\frac{d^2\mu}{d\varphi^2}+\mu=3M\mu^2+(\frac{E^2}{L^2}\frac{1}{\mu}-2\mu+5M\mu^2+3M\mu^2\text{ln}\frac{1}{\mu})d,
\end{equation}
which can also be simplified in the larger radius case, i.e. $\mu=1/r\sim0$ as
\begin{equation}
\frac{d^2\mu}{d\varphi^2}+\mu=3M\mu^2+\frac{E^2}{L^2}\frac{d}{\mu}.
\end{equation}
If we solve this equation in exact the same way used before, we obtain
\begin{equation}
\mu(\varphi) = \frac{1}{l}\text{sin}\varphi+\frac{M}{l^2}(1-\text{cos}\varphi)^2 + \frac{E^2l}{L^2}(-\varphi\text{cos}\varphi + \text{sin}\varphi\text{ln sin}\varphi)d,
\end{equation}
which can be further simplified near $\varphi=\pi$
\begin{equation}
\mu(\varphi) = \frac{1}{l}\text{sin}\varphi+\frac{M}{l^2}(1-\text{cos}\varphi)^2 + \frac{E^2l}{L^2}(-\varphi\text{cos}\varphi) d.
\end{equation}
Therefore, the angle can be obtained
\begin{equation}
\beta=\frac{4M}{l}+\frac{E^2l(4M+\pi l)}{L^2}d.
\end{equation}

\section{Conclusion and discussion}
In this paper, the perihelion precession and
deflection of light have been investigated in the $4$-dimensional general spherically symmetric spacetime. Since the perihelion precession and deflection of light can be treated respectively as the time-like and null geodesic in spacetime, the master equation of perihelion precession and deflection of light is obtained. Moreover, the Reissner-Nordstorm-AdS solution in Einstein gravity and Clifton-Barrow solution in $f(R)$ gravity are taken for the two examples as the application of this master equation, and the effects from the cosmological constant, electric charge and $f(R)$ gravity on the perihelion precession and deflection of light are investigated. We find that the electric charge can affect both the perihelion precession and deflection of light, while the cosmological constant can only affect the perihelion precession, which are consistent with the results in the previous works. In addition, after considering the case with small $d$, we easily find that the $f(R)$ gravity can also affect both the perihelion precession and deflection of light. Several discussions related to our results are in order:

(i) During investigating the effects from the electric charge, cosmological constant and $f(R)$ gravity on the perihelion precession and deflection of light, we have assumed the approximate conditions like (\ref{Condition}) and small $d$. Therefore, the results under other approximate conditions are interesting open issues.

(ii) The perihelion precession and deflection of light can contain the information of the spacetime, i.e. the cosmological constant, electric charge and small $d$. Particularly, if we just consider them in the solar system, since the angle of perihelion
precession and deflection of light in the solar system can be detected by the
experiments, thus the information like cosmological
constant and electric charge can be extracted from the angles by the experiments. Therefore, we may also extract the information of dark matter such
as its distribution of energy density $\rho(r)$ through detecting
the perihelion precession and deflection of light by the
experiments if we can first find the static solution containing the dark matter, which will be further studied in the future work.

(iii) A simple constraint on the cosmological constant. Although the
cosmological constant does not affect the deflection of light, it
can affect the perihelion precession. From our results, if the cosmological constant
indeed exists in our universe, it can also give effects in the
solar system. Therefore, we can give a simple constraint on the cosmological constant from the well known phenomenon-perihelion precession of
Mercury in the solar system. The experiment data of anomalous precession angle is
$(42.56\pm0.94)''$ per century~\citep{Clemence-1947}. Therefore, the
theoretical result should be in $(42.56\pm0.94)''$. Considering the simple case $Q=0$ in (\ref{M3_7}), and recovering
the constants $G$ and $c$, Eq.(\ref{M3_7}) becomes
\begin{equation}
\Delta \varphi=2\pi(\frac{3M^2G^2}{L^2c^2}-\frac{\Lambda
L^6}{2M^4G^3}).
\end{equation}
Since the theoretical result of the first term is $43.03''$ per
century, thus the possible contribution from cosmological constant
$\frac{\pi\Lambda L^6}{M^4G^3}$ must be less than $1.41''$ per
century, where we just consider the positive cosmological constant which can accelerate our universe. After inserting the constants, the mass of sun $M=1.989
\times 10^{30}Kg$ and the angular momentum of unit mass of Mercury
$L=2.72 \times 10^{16} m^2s^{-1}$, we can constrain the
cosmological constant that $\Lambda < 5.89 \times 10^{-11}Kg/m^3$,
which is consistent with observation data in our universe $\Lambda =
1.9 \times 10^{-25} \text{Kg/m}^3$.

(iv) The subtlety in the deflection of light in the solar system. Note that, there is
a gap between the experiment data $(1.61\pm0.4)^{\prime\prime}$ and
theoretical result $1.75^{\prime\prime}$ in Schwarzschild
spacetime~~\citep{Dyson-Eddington-Davidson-1920,Kennefick:2009}. From the result in
(\ref{QstarD}), one may consider that this difference may be
from the sun's possible electric charge effect. However,
we will give a simple proof in the following that it is not true.
After recovering the constants, the angel of starlight deflection is
\begin{equation}
\beta=\frac{4MG}{c^2l}-\frac{3\pi Q^2G}{4l^2c^4},
\end{equation}
where we have use the Gauss unit of Q, i.e. $kg^{1/2}m^{3/2}s^{-1}$.
From which, we can find that the charge could indeed make the angel
smaller, and hence make the gap smaller. If the charge could make
the angle smaller $1^{\prime\prime}$, i.e. $\frac{3\pi
Q^2G}{4l^2c^4}\times 180/ \pi \times60\times60=1$, we can obtain
the sun with possible charge $Q = 1.1 \times 10^{28}kg^{1/2}m^{3/2}s^{-1}$.
However, note that the sun's charge can also affect the perihelion
precession of Mercury in (\ref{M3_7}), which can be recovered
\begin{equation}
\Delta \varphi=2\pi(\frac{3M^2G^2}{L^2c^2}-\frac{Q^2G}{2L^2c^2}).
\end{equation}
where we have neglected the effect of cosmological constant since the observation data in our universe $\Lambda =
1.9 \times 10^{-25} \text{Kg/m}^3$ is very small. Therefore, after the simple calculation, we can obtain that the sun's charge effect on the angle of perihelion precession of Mercury will be $\frac{ \pi Q^2G}{L^2c^2}=0.038$, which is larger than the first term $\frac{6 \pi M^2G^2}{L^2c^2}=4.97\times10^{-7}$. This is obviously opposite to the experiment observation. Thus the gap between the experiment data and
theoretical result for the starlight deflection in solar system could not completely come
from the sun's electric charge, and the more suitable explanation of this difference will be still an open issue.

\section*{Acknowledgments}
This work is supported by National Natural Science Foundation of
China (NSFC) under grant No.11105004 and Shanghai Key Laboratory of
Particle Physics and Cosmology under grant No.11DZ2260700, and
partially by grants from NSFC (No. 10821504, No. 10975168 and No.11035008) and the Ministry of Science and Technology of China under
Grant No. 2010CB833004. Prof. Hongsheng Zhang is supported by the Program for Professor of Special Appointment (Eastern Scholar) at Shanghai Institutions of Higher Learning, and National Natural Science Foundation of China under Grant Nos. 11075106, 11275128.




\end{document}